# A feasibility study on filtering low-accessibility web pages considering color vision deficiency

**Ryota MIZUTANI†, Shiori NAKAYAMA†, and Masateru TSUNODA†,††**

**SUMMARY** Recently, the importance of universal design has increased. Color universal design (CUD) is one type of universal design that takes people with color vision deficiency (CVD) into consideration. Websites are important media for providing various types of information and functions. Therefore, it is essential to enhance the accessibility of web pages by incorporating CUD principles. The goal of our study is to help improve the accessibility of web pages. Our approach is to automatically filter low-accessibility web pages. To evaluate the feasibility of this approach, we conducted an experiment using 21 web pages. The prediction model identified low-accessibility pages with reasonable accuracy, achieving a maximum AUC of 0.76.

## 1. Introduction

Recently, providing reasonable accommodations for people with disabilities has become legally required in Japan, and the importance of universal design has increased. Universal design means that products and services are designed to be usable by people regardless of their age or disabilities. Color universal design (CUD) is one type of universal design. Some people have color vision deficiency (CVD), which means their visual perception differs from that of the majority, and it can be difficult for them to distinguish certain colors. CUD takes such people into consideration.
Websites are very important media for providing various types of information and functions. Therefore, it is important to enhance the accessibility of web pages by considering CUD, and previous studies have focused on the accessibility of web pages for people with CUD [4][6][7].

Generally, to improve web accessibility, the CUD of web pages is evaluated, and the color schemes of pages with low accessibility are improved based on the evaluation. Web sites often consist of many pages; therefore, automated filtering of low-accessibility pages is an effective way to reduce the cost of CUD improvements compared with manual filtering.
The goal of our study is to automatically identify low-accessibility web pages. By inputting an image of a web page into a prediction model, the model classifies whether the page is accessible for people with CVD. The basic idea of our study is akin to that of Study [9], which classifies problematic images in the biology literature, considering dichromatic vision. Compared with that study, the major contributions of our work are as follows:

Basic idea of our study is akin to study [9], which classifies problematic images in the biology literature. Compared with the study, major contributions of our study are follows:

- Propose to filter low-accessibility web pages automatically
- Evaluate the feasibility of the automatic filtering
- Consider protanopia and tritanopia in addition to dichromatic vision

## 2. Related work

Contrast of web pages is also related to accessibility, and the Web Content Accessibility Guidelines (WCAG) establish preferable contrast levels for web pages. Based on WCAG, there are tools that automatically evaluate accessibility according to the contrast of web pages, and Kumar et al. [5] evaluated such tools. However, these tools do not consider the visual perception of people with color vision deficiency (CVD), and it is difficult to accurately identify web pages with low accessibility for people with CVD.

Considering people with CVD, Meenakshi et al. [6] proposed a web accessibility assessment process. In this process, tools are used to evaluate web pages, and people assess accessibility based on the tools' evaluations and manual observations. Although the tools' evaluations include the contrast of web pages, they do not include other color-related issues. That is, using this process, we cannot automatically identify web pages with low accessibility for people with CVD.

To assess the accessibility of web pages, there is an approach in which web pages are transformed into images whose colors are adjusted to the vision of people with CVD. Jamil et al. [4] applied this approach and analyzed the relationship between the aesthetics and functionality of user interfaces of web pages for people with CVD. To conduct the analysis, non-CVD participants rated the interfaces using these images on a four-grade scale. They concluded that there is a positive relationship between the functionality and aesthetics of web pages. However, this study did not propose any automated process to identify web pages with low aesthetics and functionality.

† Faculty of Informatics, Kindai University, Higashiosaka-shi, 577-8502 Japan.
†† Cyber Informatics Research Institute, Kindai University, Higashiosaka-shi, 577-8502 Japan.

Sajek et al. [7] defined a color difference metric to analyze the accessibility of web pages for people with CVD. Although this metric was used to quantify color differences, it was not employed to analyze their quantitative relationship with accessibility. Therefore, it is not clear whether the metric is effective in identifying low-accessibility pages.

Zhuang et al. [10] proposed a metric that evaluates how accessible an image is for people with CVD. This metric assumes that it is applied to ordinary images (i.e., not documents) and focuses on the loss of edges in images. However, the factors affecting the accessibility of web pages are not limited to edges. For instance, when types of information are signified by color, accessibility could be low even if edge information is not lost. Additionally, the validation of the metric was not conducted through subjective experiments. Therefore, it is unclear whether the metric can be used to identify web pages with low accessibility.

In another field, there is a study that automatically identifies problematic images considering people with CVD. Stevens et al. [9] proposed a method to build a model that classifies problematic images in the biology literature. The basic idea of our method is the same as that of their study. However, their study considers only dichromatism, and their method is limited to images in the biology literature. In our experiment, we compared the accuracy of their model with that of our method. One of our contributions is to show that web pages with low accessibility for people with CVD can be automatically identified, as stated in Section 1.

There is a study that converts images into others with enhanced accessibility for people with CVD [8]. However, such conversion does not guarantee the preservation of image aesthetics, and, as mentioned above, the aesthetics of web pages are related to functionality [4]. This suggests that the conversion is not always effective in addressing issues of web page accessibility. Additionally, our method is expected to trigger revisions of web design policies for each website, which would fundamentally improve accessibility.

## 3. Classification of low-accessibility pages

To automatically identify low-accessibility web pages, screenshots of web pages are input into a machine learning model, and the model classifies the images as either low accessibility or not. Overall, the procedure for building the model is the same as in study [9], while the algorithm used is different from that study.

**Simulator**: In this procedure, a vision simulator tool for people with CVD is used to collect values of dependent and independent variables for the model. In this study, we used a tool [1] that can simulate type D (dichromatism), P (protanopia), and T (tritanopia) vision types, respectively. The simulation is based on the study by Brettel et al. [2].

**Color features**: The classification model is based on the assumption that color perception differs significantly between people with CVD and non-CVD on low-accessibility pages. Based on this assumption, color features of web pages are used as independent variables in the model. These color features are derived from RGB and HSV model respectively. RGB model represents color using red, green, and blue, while the HSV model represents it using hue, saturation, and brightness (value). Note that the HSV model was not used in study [9]; therefore, this is one of the differences between our study and previous work.

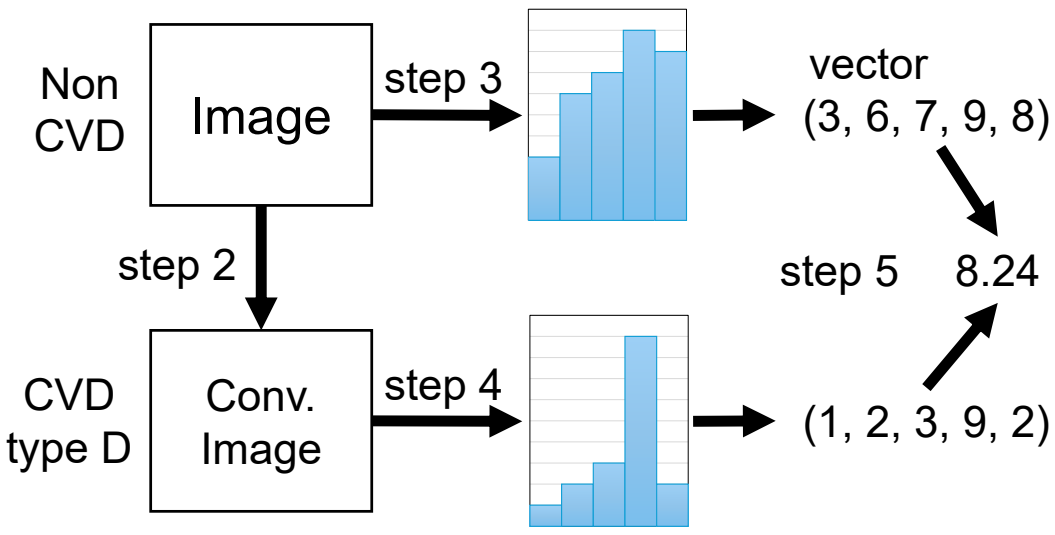

**Fig. 1.** Procedure to calculate difference of color histograms

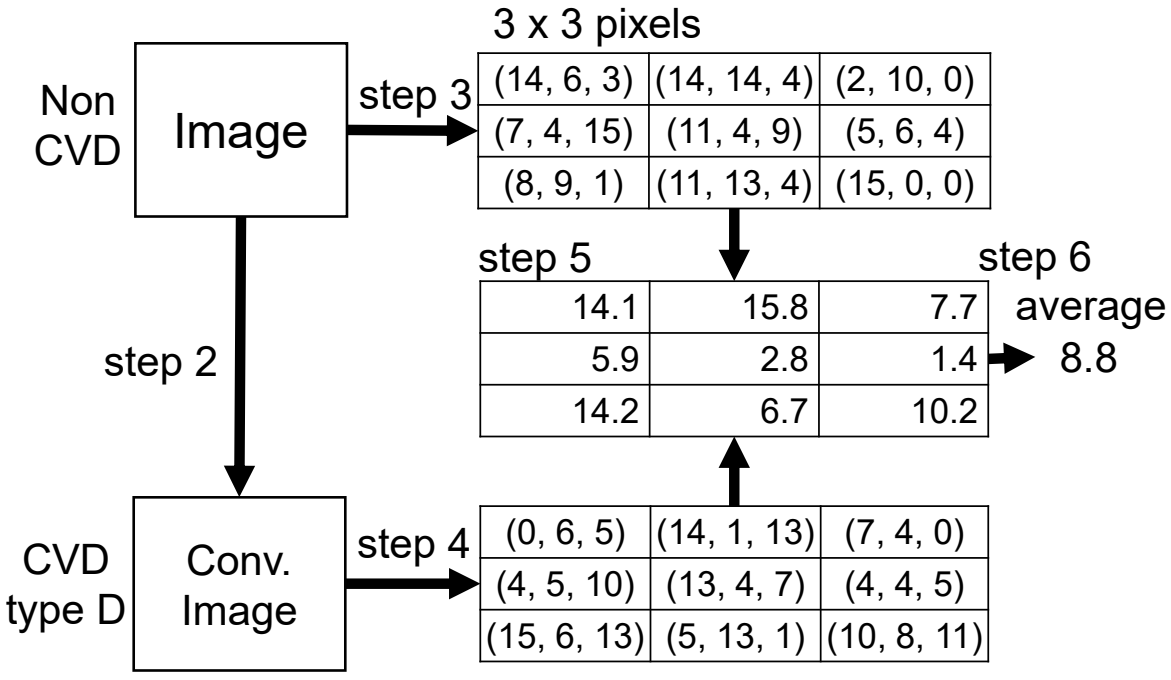

**Fig. 2.** Procedure to calculate color difference on each pixel

**Difference of color histograms**: Based on the differences in color histograms in the RGB and HSV models, our procedure generates color features (i.e., independent variables) following the procedure below. The procedure is based on study [3]; however, the goal of that study was to retrieve similar images, and it used only the RGB model, not the HSV model.

1. Obtain a screenshot of the target web page.
2. Convert the screenshot based on type D perception using the simulator.
3. Generate frequency distribution tables (histograms) for red, blue, and yellow, respectively, using the screenshot from step 1.
4. Similarly, using the screenshot from step 2, generate three frequency distribution tables.
5. Treat the tables from steps 3 and 4 as vectors, and calculate the Euclidean distances for the red, blue, and yellow vectors, respectively.

The above procedure is an example in which the RGB model and type D perception are used. Fig. 1 illustrates this procedure. Similarly, the procedure uses the HSV model and type P and T, respectively, to generate color features. The number of classes in the histograms is 256, because color

**Table 1.** Example of generated data from a screenshot of a web page

| type | eval. | eval.<3 | hist. R | hist. G | hist. B | hist. H | hist. S | hist. V | dist. RGB | dist. HSV |
|---|---|---|---|---|---|---|---|---|---|---|
| D | 2 | 1 | 12.1 | 12.1 | 12.0 | 12.9 | 12.1 | 12.1 | 16.2 | 19.4 |
| P | 3 | 0 | 12.1 | 12.1 | 12.0 | 12.7 | 12.1 | 12.1 | 18.8 | 22.7 |
| T | 4 | 0 | 12.0 | 11.9 | 11.9 | 13.2 | 12.7 | 12.0 | 9.6 | 15.6 |

test data

| web page | type | eval. < 3 | ... |
|---|---|---|---|
| α | D | 1 | ... |
| α | P | 1 | ... |
| α | T | 0 | ... |
| β | D | 1 | ... |
| β | P | 0 | ... |
| β | T | 1 | ... |
| γ | D | 1 | ... |

test data

| web page | type | eval. < 3 | ... |
|---|---|---|---|
| α | D | 1 | ... |
| α | P | 1 | ... |
| α | T | 0 | ... |
| β | D | 1 | ... |
| β | P | 0 | ... |
| β | T | 1 | ... |
| γ | D | 1 | ... |

**Fig. 3.** Example of alternative leave-one-out cross validation

values in programming languages such as Python are often represented with 256 levels. The frequency of each class denotes the number of pixels whose corresponding color value (e.g., red) falls into that class. We define these variables as hist. R, hist. G, hist. B, hist. H, hist. S, and hist. V.

**Color difference at each pixel**: Based on study [9], the color difference at each pixel is calculated to derive color features (i.e., independent variables). The following procedure describes this calculation. Note that, compared with study [9], the procedure is simplified to apply it easy. Additionally, that study used only the RGB model.

1. Obtain a screenshot of the target web page.
2. Convert the screenshot based on type D perception using the simulator.
3. For each pixel in the screenshot from step 1, construct a three-dimensional vector based on the red, blue, and green values.
4. Similarly, for each pixel in the screenshot from step 2, construct a three-dimensional vector.
5. Calculate the Euclidean distance between the vectors from steps 3 and 4 for each pixel.
6. Compute the average Euclidean distance from step 5.

The above procedure is an example using the RGB model. Fig. 2 illustrates this procedure. Similarly, the procedure uses the HSV model and types D, P, and T, respectively, to generate color features. Note that, in some programming libraries such as OpenCV, the range of hue is 180. In such cases, the range is aligned with that of saturation and brightness (e.g., 256). Additionally, the color difference between hue values 0 and 179 is very small, because the next value after 179 is 0, similar to a clock. To calculate the difference in hue, we applied the following equation:

$$d = \min(|a - b|, 180 - |a - b|) \quad (1)$$

In this equation, $d$ represents the difference in hue, and $a$ and $b$ represent the hue values of the pixels. We define these variables as dist. RGB and dist. HSV.

**Accessibility evaluation**: The accessibility of web pages (i.e., the dependent variable) is evaluated using the following procedure. As in previous studies [5][8], people with non-CVD evaluate accessibility.

1. Obtain a screenshot of the target web page.
2. Convert the screenshot based on type D perception using the simulator.
3. By comparing the screenshots from steps 1 and 2, people with non-CVD evaluate the accessibility of the screenshot from step 2 using a five-point scale (1: low, ..., 5: high).

The above procedure is an example when type D perception is used. Similarly, the procedure uses types P and T, respectively. To build the classification model, we regard cases as positive when the evaluation is less than 3, and the dependent variable is defined according to this threshold. We define these variables as eval. and eval.<3.

Table 1 shows an example of the data generated by the above procedures using a single snapshot. From this snapshot, three data points are obtained, each consisting of one dependent variable and eight independent variables.

## 4. Experiment

**Dataset**: To evaluate the accuracy of the model that identifies low-accessibility web pages, we conducted an experiment. We selected 21 web pages from 21 different Japanese websites, such as e-commerce sites and local government sites. Using the CVD simulation tool described in Section 3, we evaluated the accessibility of these pages. In study [9], images were assessed by the authors, and we followed the same approach. In the evaluation, we considered the visibility of objects and information loss by comparing the simulated pages with the original pages.

As shown in Table 1, three data points are generated from each web page by the procedure. Therefore, we generated 63 data points for the experiment. The number of low-accessibility pages (i.e., eval.<3 = 1) was 23 out of the 63 data points.

To build and evaluate the classification models, we split the dataset into training and test datasets. For this split, we applied an alternative leave-one-out cross-validation method. Each web page has three data points, and these data points were not divided between the training and test sets to avoid the risk of data leakage. For example, as shown in Fig. 3, data points related to web page α are treated as test data, while the others are treated as training data in the first iteration. In the next iteration, data points related to web page β are treated as test data.

**Classification models**: As classification models, we used k-nearest neighbors (k-NN), naïve Bayes, neural networks (NN), and random forests based on preliminary analysis. As independent variables, we used the following combinations:

- **ALL**: hist. R, hist. G, hist. B., hist. H, hist. S, hist. V, dist. RGB, and dist. HSV
- **HIST**: hist. R, hist. G, hist. B., hist. H, hist. S, and hist. V
- **DIST**: dist. RGB, dist. HSV
- **RGB**: hist. R, hist. G, hist. B., and dist. RGB
- **HSV**: hist. H, hist. S, hist. V, and dist. HSV

**Table 2.** AUC of each model

| combination | k-NN | naïve Bayes | NN | random forests |
|---|---|---|---|---|
| ALL | 0.62 | 0.70 | 0.57 | 0.50 |
| HIST | 0.50 | 0.69 | 0.54 | 0.46 |
| DIST | 0.76 | 0.68 | 0.62 | 0.70 |
| RGB | 0.55 | 0.69 | 0.56 | 0.59 |
| HSV | 0.65 | 0.63 | 0.53 | 0.53 |

**Table 3.** Prediction accuracy of the baselines

| | random | model of study [9] |
|---|---|---|
| **AUC** | 0.53 | 0.60 |

**Evaluation Criterion**: As the evaluation criterion, we used AUC (area under the curve), which was also used in study [9]. As baselines, we used random classification and the published model from study [9] (https://bioapps.byu.edu/colorblind_image_tester). We repeated the cross-validation process 10 times, since some models include random factors.

**Result**: The prediction accuracy of the models and the baselines is shown in Tables 2 and 3. Overall, naïve Bayes achieved higher accuracy, and k-NN using DIST achieved the highest accuracy. Regardless of the combinations of independent variables, the accuracy of naïve Bayes was higher than that of the baseline models. Note that the model in study [9] was trained using images from the biology literature. Therefore, the result does not indicate that the method of study [9] is not effective for identifying low-accessibility pages; rather, it suggests that the existing tool is not effective without calibration.

When the DIST combination was used, the accuracy was high across all four classification methods. In contrast, for naïve Bayes, the ALL, HIST, RGB, and DIST combinations also achieved high accuracy. The dataset used in this study was small; therefore, the DIST combination may not always be the best. It would be preferable to test all combinations when building a classification model.

**Threats to validity**: Comparing pixel-wise color differences and histogram differences has been used in previous studies [3][9]; therefore, using these measures has a certain degree of validity. The dataset used in this study was small, and when our approach is applied to a wider variety of web pages (i.e., increasing the amount of test data), the accuracy of the model may decrease. However, by increasing the amount of training data, it would be possible to use more complex independent variables and classification methods such as XGBoost. This is expected to improve prediction accuracy. At least, the experimental results suggest that classifying low-accessibility web pages is feasible.

## 5. Conclusion

The goal of our study is to identify low-accessibility web pages for people with color vision deficiency. To achieve this goal, we experimentally evaluated the feasibility of models that classify low-accessibility web pages. In the model, we considered vision types of dichromacy, protanopia, and tritanopia. To construct the independent variables of the model, we focused on differences in color histograms and pixel-wise color differences, considering both RGB and HSV models.

In the experiment, we collected 21 web pages and constructed a dataset. It included 63 data points by considering the accessibility of three vision types, respectively. The maximum AUC of the model was 0.76, which was higher than the baselines (i.e., random classification and a model based on a previous study). The results suggest that our approach can achieve reasonable accuracy in identifying low-accessibility pages.

## Acknowledgments

This research is partially supported by the Japan Society for the Promotion of Science [Grants-in-Aid for Scientific Research (C) (No. 26K14789).